\documentclass{PoS}

\title{Brief Report on the Extended Linear Sigma Model}

\ShortTitle{Brief Report on the Extended Linear Sigma Model}

\author{\speaker{Anja Habersetzer}%
         \\
        Institut f{\"u}r Theoretische Physik, Goethe-Universität, Frankfurt am Main\\
        E-mail: \email{habersetzer@th.physik.uni-frankfurt.de}}

\abstract{We present a chiral effective model which exhibits all the known symmetry features of the QCD Lagrangian. The extended Linear Sigma Model (eLSM) includes (pseudo-)scalar and (axial-) vector mesons, glueballs, and baryons. It has proven to be a useful tool to understand the nature of the low-energy resonances. It can be used to describe the masses and decay widths of the known low-energy mesons and to disentangle the complicated picture of resonances in the low-energy range. We can answer some of the open questions in the low-energy region, e.g. the resonance \(f_0(1500) \) is a predominantly glueball state and the resonance \(f_0(1370)\) is most likely the chiral partner of the pion. We could also find that the resonance \(N(1650)\) is favoured as the chiral partner of the nucleon \(N(939)\). In addition the temperature dependence of the chiral condensate and the glueball condensate can be used to study the phase diagram at nonzero temperatures and 
densities. At nonzero density we achieve nuclear matter saturation.}

\FullConference{Xth Quark Confinement and the Hadron Spectrum\\
                 8.-12. October 2012\\
                 TUM Campus Garching, Munich, Germany}

\usepackage{mathrsfs}
\usepackage{amsmath}
\usepackage[usenames,dvipsnames]{xcolor}
\usepackage{empheq}
\usepackage{feynmp}

\begin{document}
Phenomenological effective models are motivated by Quantum Chromodynamics. They are a useful tool to describe the hadronic world in the low-energy region. The running coupling \(\alpha_S\) is large at \(\sim 1 \) GeV and quarks and gluons are confined to colorless bound states. A microscopic description of these bound states is then replaced by a phenomenological description of the hadrons and their interactions. 

The bound states of the eLSM are the (pseudo-)scalar and (axial-)vector mesons, the scalar-isoscalar glueball and the nucleon together with its chiral partner.
In the limit of vanishing quark masses the QCD Lagrangian is invariant under chiral transformations. This \(U(N_f)_L \times U(N_f)_R\) flavor symmetry and dilatation invariance, together with their explicit, spontaneous and anomalous symmetry breaking patterns, are the guiding principles in the construction of the eLSM. Dilatation invariance (which is broken explicitly in the gluonic sector and by the nonzero quark masses) ensures that the total number of allowed terms in the eLSM is finite. The ideas how the symmetry features of the QCD Lagrangian are modelled by the eLSM in the meson sector with glueball and in the baryon sector will be briefly outlined in the following. For details the reader is referred to Refs. \cite{Parganlija:2010fz,Parganlija,Janowski:2011gt,Gallas,Gallas02,habersetzer} and the references therein. 

 The core part of this model is the chirally invariant mesonic Lagrangian, first  presented for \(N_f=2\) in \cite{Parganlija:2010fz} and then extended to \(N_f=3\) in \cite{Parganlija}. It contains the low-energy (pseudo-)scalar and (axial-)vector isospin multiplets. The \(U(N_f)_L \times U(N_f)_R\) symmetry is broken explicitly and spon\-taneously. It is explicitly broken to \(U(2)_V\) by (i) a term which corresponds to the \(U(1)_A\) anomaly of the QCD Lagrangian, and (ii) a term that accounts for the explicit breaking of chiral symmetry through the quark masses. In the case of isospin breaking, \(m_u \neq m_d\), the remaining symmetry is \(U(1)_V\).  The gluon condensate that arises via the trace anomaly of QCD generates the spontaneous breaking of chiral symmetry. The chiral condensate then explains the mass difference between the chiral partners (e.g. \(a_1\) and \(\rho\) or \(N\) and \(N^*\)). 
\begin{figure}[hb]
\includegraphics[width=0.45\textwidth]{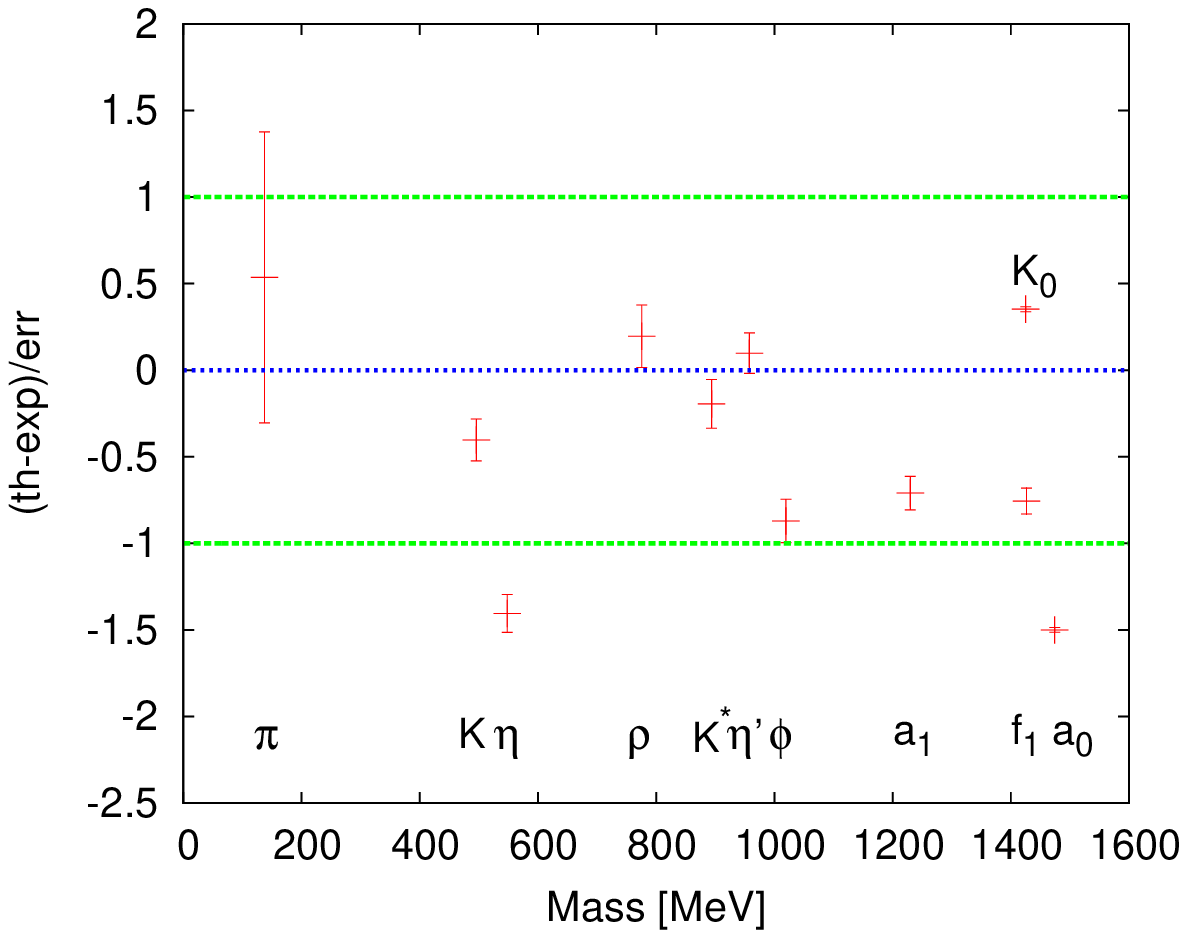}
\includegraphics[width=0.45\textwidth]{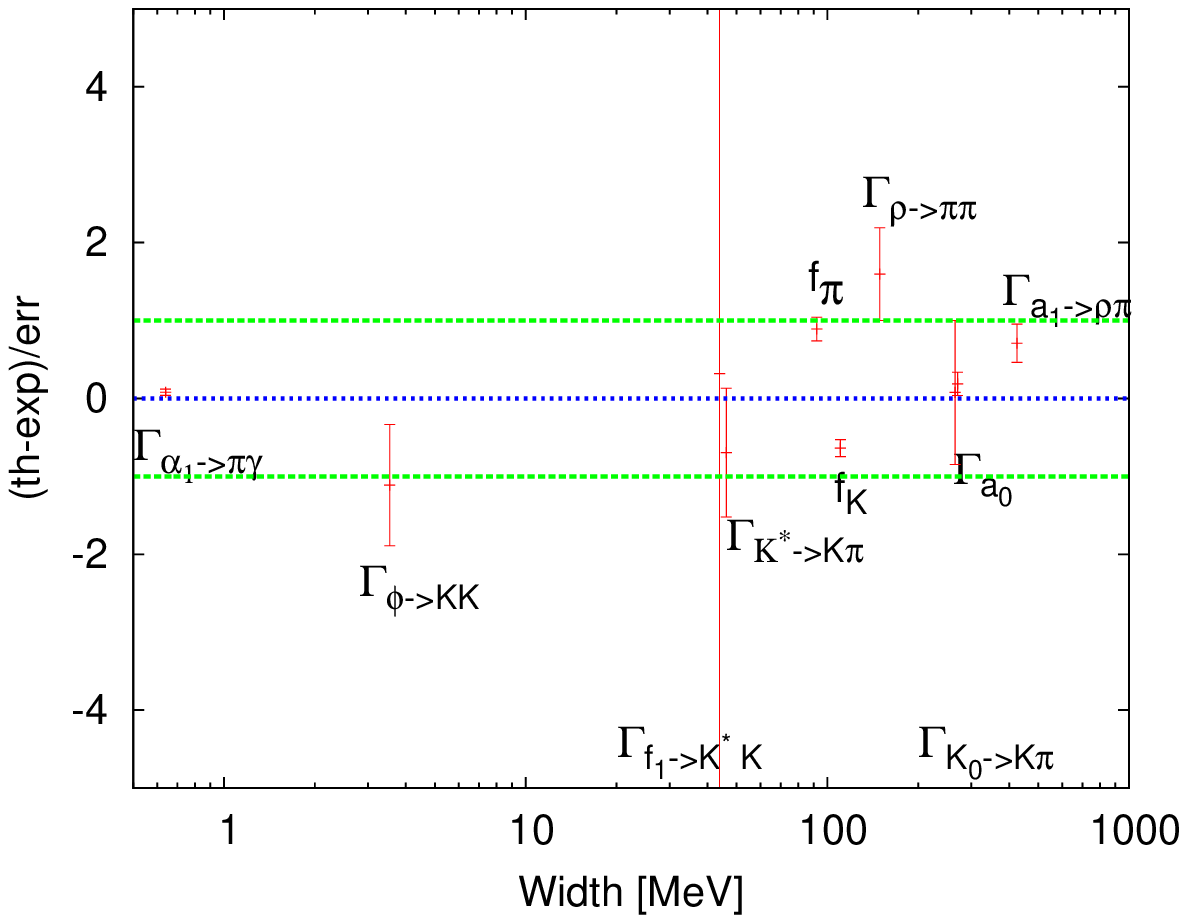}
\caption{Comparison of results from the \(N_f=3\) meson sector to the experimental data.}\label{ParganlijaResults}
\end{figure}
Various masses and decay widths in the \(N_f=3\) meson sector have been calculated in Ref. \cite{Parganlija}. The comparison of the results with experimental data is shown in Figure \ref{ParganlijaResults}.
 
 The fluctuation of the dilaton, the glueball, was implemented into the model for \(N_f=2\) in Ref. \cite{Janowski:2011gt}. It was the first time that the glueball was included into a linear sigma model with (axial-)\-vector mesons which allowed for calculating decays of the glueball into vector mesons. It was found that the \(J^{PC}=0^{++}\) resonance \(f_0(1370)\) as a predominantly \(\bar{q}q \) state reproduces the experimental data and confirms the assumption that \(f_0(1370)\) is the chiral partner of the pion.  The resonance \(f_0(1500)\) as a predominantly glueball state yields the best result for the decays of \(f_0(1500)\) in the \(\pi\pi\), \(\eta\eta\), and \(KK\) channels.  The scenario in which \(f_0(600)\) is the chiral partner turns out to be strongly disfavoured. The role of the (axial-)vector mesons for these results is crucial.

 Chiral symmetry for the nucleon is realized by the so-called mirror assignment of the bare nucleon fields (\(\Psi_1 \ , \ \! \Psi_2\)) \cite{Gallas}. In the mirror assignment the physical fields \(N\) and \(N^*\) are truly chiral partners that belong to the same multiplet and the behavior of the bare fields under chiral transformations allows for the inclusion of a bare nucleon mass term without violating chiral symmetry. The bare nucleon mass is interpreted as being generated by the gluon and/or the tetraquark con\-den\-sate. The physical nucleon mass is then interpreted as a result of these two condensates and a contribution from the chiral condensate. The mirror assignment yields the correct nuclear matter behavior for quantities such as nuclear matter saturation \cite{Gallas02}. The pion-nucleon scattering lengths have been calculated within this approach in Ref. \cite{Gallas}. With the resonance \(N(1650)\) as chiral partner of the nucleon \(N(939)\) the scattering lengths are in good agreement with 
the data (e.g. \(a_0^{(-) eLSM} = (5.90 \pm 0.46)\cdot10^{-4}\text{ MeV}^{-1}\ ,\ \! a_0^{(-) exp.} = (6.
4 \pm 0.1)\cdot10^{-4}\text{ MeV}^{-1}\)). It was found that 
the role of the (axial-)vector mesons is important to reproduce experimental results in the vacuum as well as at nonzero densities. 
 
The chiral symmetry of the eLSM can also be used to include electroweak interactions. This was done by performing a \(U(1)_Y\times SU(2)_L\) gauge transformation of the (pseudo-)scalar fields. The vector channel in the decay of the \(\tau\) lepton has been calculated in \cite{habersetzer} and the work on the axial-vector channel is in progress.

The eLSM is a good phenomenological tool to describe low-energy QCD. The extension of the mesonic part of the eLSM to \(N_f=4\) is in progress as well as the extension of the gluonic scalar-isoscalar part to \(N_f=3\). The next step will be the extension of the baryonic sector to \(N_f=3\) and the inclusion of the \(\Delta\) resonance which yields also an important contribution to the dilepton decay rate. Then e.g. also the in-medium modifications of the \(\rho\) resonance can be studied.


\begin{thebibliography}{99}
\bibitem{Parganlija:2010fz}
  D.~Parganlija, F.~Giacosa and D.~H.~Rischke,
  Phys.\ Rev.\ D {\bf 82} (2010) 054024
  [{\tt arXiv:1003.4934 [hep-ph]}].
\bibitem{Parganlija}
  D.~Parganlija, P.~Kovacs, G.~Wolf, F.~Giacosa and D.~H.~Rischke,
 [{\tt arXiv:1208.0585 [hep-ph].}]
\bibitem{Janowski:2011gt}
  S.~Janowski, D.~Parganlija, F.~Giacosa and D.~H.~Rischke,
  Phys.\ Rev.\ D {\bf 84} (2011) 054007
  [arXiv:1103.3238 [hep-ph]].
\bibitem{Gallas}
  S.~Gallas, F.~Giacosa and D.~H.~Rischke,
  Phys.\ Rev.\ D {\bf 82} (2010) 014004
  [{\tt arXiv:0907.5084 [hep-ph]}].
  \bibitem{Gallas02} S.~Gallas, F.~Giacosa and G.~Pagliara,
  Nucl.\ Phys.\ A {\bf 872} (2011) 13
  [{\tt arXiv:1105.5003 [hep-ph]}].
\bibitem{habersetzer}A. Habersetzer and F. Giacosa,  
{\emph Acta Physica Polonica} B Proc. Suppl. {\bf 5/4}, 1077 (2012).
\end{thebibliography}
\end{document}